\documentclass[aps,prl,twocolumn]{revtex4}
\usepackage{graphicx}
\usepackage{amsmath}
\usepackage{amsfonts}
\usepackage{color}
\graphicspath{{./density/}}

\begin{document}

\title{Computation of the unifying thread in high temperature superconductors from first principles quantum Monte Carlo}

\author{Awadhesh Narayan}
\altaffiliation{Present address: Materials Theory, ETH Zurich, Wolfgang-Pauli-Strasse 27, CH 8093 Zurich, Switzerland.}

\author{Brian Busemeyer}

\author{Lucas K. Wagner}
\affiliation{Department of Physics, University of Illinois at Urbana-Champaign, Urbana, Illinois, USA.}

\date{\today}

\begin{abstract}
It has long been a challenge to describe the origin of unconventional superconductivity. 
The two known examples with high T$_c$, based on iron and copper, have very different electronic structures, while other materials with similar electronic structure may not show superconductivity at all.
In this paper, the authors show that by using high accuracy diffusion Monte Carlo calculations, the unconventional superconductors of both high T$_c$ types form a cluster at intermediate spin-charge coupling. 
The spin-charge coupling may serve as a normal state marker for unconventional superconductivity, and provides evidence that unconventional superconductivity is due to interaction of charge with local spins in materials.
\end{abstract}

\maketitle

\textit{Introduction:} 
Since its discovery in the cuprates, high temperature superconductivity has remained a fascinating puzzle in condensed matter physics~\cite{bednorz1986possible}. 
In these systems, while phonons may or may not be important, it seems clear that other causes are necessary to explain their high critical temperatures. 
For much of the time that high temperature superconductivity has been known, there has been one class of materials that exhibited high temperature superconductivity---the cuprates. 
This is in stark contrast to electron-phonon superconductors, for which there are many varied examples. 
This solitary class presents a real challenge to finding the root cause of high temperature superconductivity. 
It is very difficult to disentangle whether an observation on the cuprates is important for the superconductivity or whether it is simply specific to that material class.

In the last decade a new class of unconventional superconductors has emerged in compounds containing magnetic element iron~\cite{kamihara2006iron,kamihara2008iron}.
Among the variety of known iron based superconductors, the BaFe$_2$As$_2$ material (122) in the ThCr$_2$Si$_2$ structure has been particularly well studied, since stoichiometric, large crystals can be synthesized~\cite{rotter2008superconductivity}.
This crystal structure is prolific; in particular, it is possible to synthesize the same structure with transition metal elements ranging from Cr to Cu. 
However, the iron 122 materials obtain a superconducting transition temperature of up to 38 K~\cite{rotter2008superconductivity2}, which stands in stark contrast to other materials in the same structure that differ only by the transition metal element and are mostly not superconducting. 
It is still unknown why in this large class of materials, the element iron is the one that attains superconductivity at high temperatures. 
The discovery of the iron-based superconductors offers a unique opportunity to find commonalities between these two different classes of high temperature superconductors.

There have been many proposals for the origin of unconventional superconductivity in both the iron-based superconductors and the cuprate superconductors~\cite{mazin2008unconventional,wang2009antiferromagnetic,anderson1987resonating,bickers1987cdw}.
While many statements on this subject are controversial, it is clear that in both the iron-based and cuprate superconductors, the superconductivity arises near the collapse of magnetic order~\cite{lumsden2010magnetism,paglione2010high}.
A magnetically mediated origin of superconducting pairing has been proposed and a better understanding of magnetic properties is widely expected to shed light on the nature of their unconventional superconductivity~\cite{scalapino2012common}.
It is thus well worth studying how magnetism and its interaction with charge excitations varies in superconducting and similar non-superconducting materials.
In particular, can we disentangle what is common between the iron based superconductors and cuprates, and different from the materials that don't superconduct?

One of the main impediments to understanding magnetism in these materials is that they are very challenging to describe theoretically, owing to the presence of both itinerant charge carriers and large localized magnetic moments~\cite{dai2012magnetism}. 
In particular, it has been a major goal in recent years to accurately describe their magnetism in a predictive manner~\cite{lebegue2007electronic,singh2008density,boeri2008lafeaso,skornyakov2009classification,yin2010unified,yin2011magnetism,yin2011kinetic,werner2012satellites,kutepov2010self,tomczak2012many}.
In early days of this field, much work using density functional theory was carried out on these systems~\cite{lebegue2007electronic,singh2008density,boeri2008lafeaso}. 
However, it was soon realized that conventionally used density functional theory does not describe the spectrum and magnetic properties of these materials accurately~\cite{yin2011magnetism,liu2008fermi,maletz2014unusual}.
There have also been several useful attempts to reproduce properties of these materials using dynamical mean field theory~\cite{skornyakov2009classification,yin2010unified,yin2011magnetism,yin2011kinetic,werner2012satellites}, as well as GW~\cite{kutepov2010self,tomczak2012many} calculations. 
A Fermi surface nesting picture has been invoked to explain superconductivity in these materials~\cite{mazin2010superconductivity}.
However, recent experiments have raised questions over such an explanation and shown that superconductivity of iron compounds does not correlate with Fermi surface topology~\cite{ye2014extraordinary}.
Therefore, the description of magnetism, as well as its interplay with superconductivity in these materials remains an open challenge.

\begin{figure}
\begin{center}
  \includegraphics[scale=1.0]{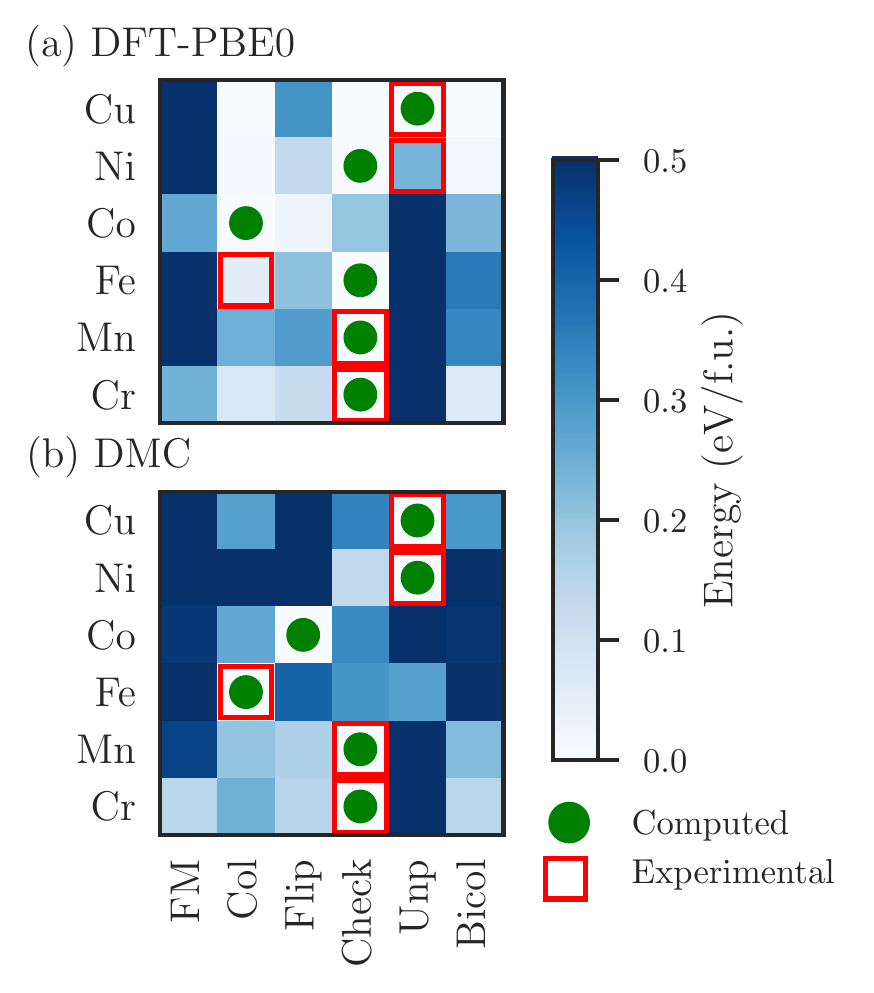}
  \caption{\textbf{High accuracy using quantum Monte Carlo.} (a) DFT-PBE0, and (b) DMC energies for different magnetic states for all the 122 compounds. 
  Blue (darker) colors represent higher energies, while white (lighter) shades indicate lower energy. 
  Red squares show the experimentally observed ground state (see text for references), while the green circles indicate the calculated ground state. 
  The scale maximum at 0.5 eV/f.u. for clarity of comparison. }  \label{energy}
\end{center}
\end{figure}

In this article, we attempt to find a link between computed material properties and high temperature superconductivity.
By using the accurate diffusion Monte Carlo technique, we can correctly describe the magnetic ground states of these materials, which signals a new achievement in the accuracy possible using \textit{ab initio} techniques on these materials. 
A comparison with more traditionally used density functional theory calculations shows the importance of including explicit correlations in describing the electronic structure of these materials.
We introduce a spin-charge coupling descriptor that links the cuprates and the iron-based superonconductors, and appears to be necessary but not sufficient for high temperature superconductivity.
Our proposed descriptor could serve as a much-sought-for normal state identifier for unconventional superconductors and could help boost the search for new unconventional superconducting systems.

\textit{Method and approach:} 
The central idea of our approach is to quantify the charge response to changes in magnetic order, in a material specific framework. 
To motivate the descriptor, let us consider a simple Kondo-Heisenberg model, in which fermions interact with a spin system:
\begin{align}
\hat{H}_{KH} &= \sum_{ij} t_{ij} \left( c_i^\dagger c_j + c_j^\dagger c_i \right)  \notag \\
    &+ \sum_{\alpha\beta} J_{\alpha\beta} S_\alpha \cdot S_\beta \\
    & + \sum_{\alpha i} K_{\alpha i} S_\alpha \cdot c_{i\sigma}^\dagger \sigma_{\sigma,\sigma'} c_{i\sigma'} 
\label{eqn:kondo_heisenberg}
\end{align}
The $i,j$ are combined lattice and orbital indices for the fermions, and $\alpha,\beta$ are lattice indices for the spins. 
This model describes free fermions interacting with a Heisenberg system of spins. 
Similar models have been considered in the context of cuprates~\cite{berg_pair-density-wave_2010} and iron pnictides~\cite{scalapino2012common}.

It is a challenge to compute whether a model like Eqn.~\ref{eqn:kondo_heisenberg} is applicable to a given material. 
First, the materials are often strongly correlated, which means that standard first principles techniques struggle to produce accurate results. 
We address this challenge by using highly accurate fixed node Diffusion Monte Carlo calculations (DMC). 
Second, Eqn.~\ref{eqn:kondo_heisenberg} is vague; it offers many possibilities for different lattices and different orbitals. 
In fact, the Kondo-Heisenberg model may not quantitatively apply to a given material. 
We address this second challenge by identifying {\it universal behavior} of all models like Eqn.~\ref{eqn:kondo_heisenberg} that can be evaluated in modern DMC calculations. 
It should be emphasized that we are using the Kondo-Heisenberg model only as an inspiration; the materials may or may not be quantitatively described by this model.

\begin{figure}
\begin{center}
\begin{tabular}{cccc}
BaFe$_2$As$_2$ \\
\includegraphics{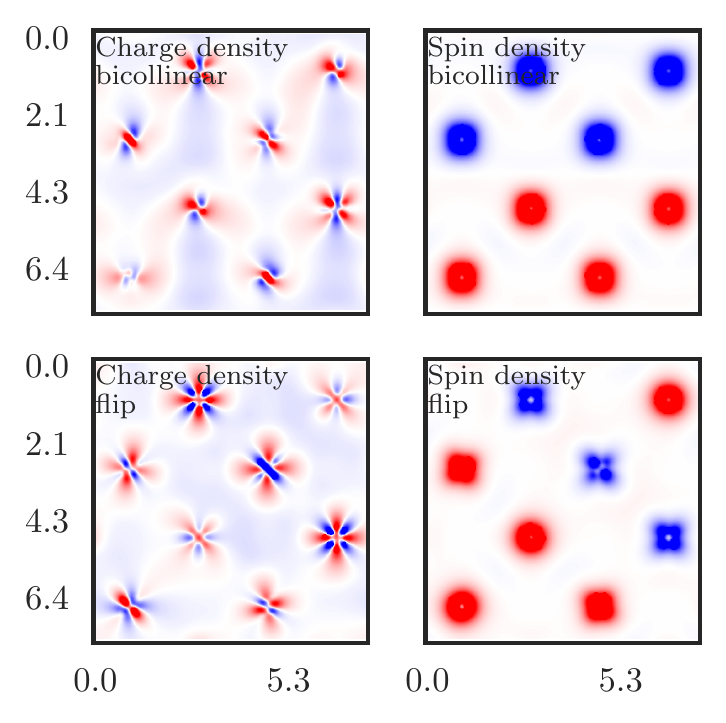} \\
Ca$_2$CuO$_2$Cl$_2$ (1/8 doping) \\
\includegraphics{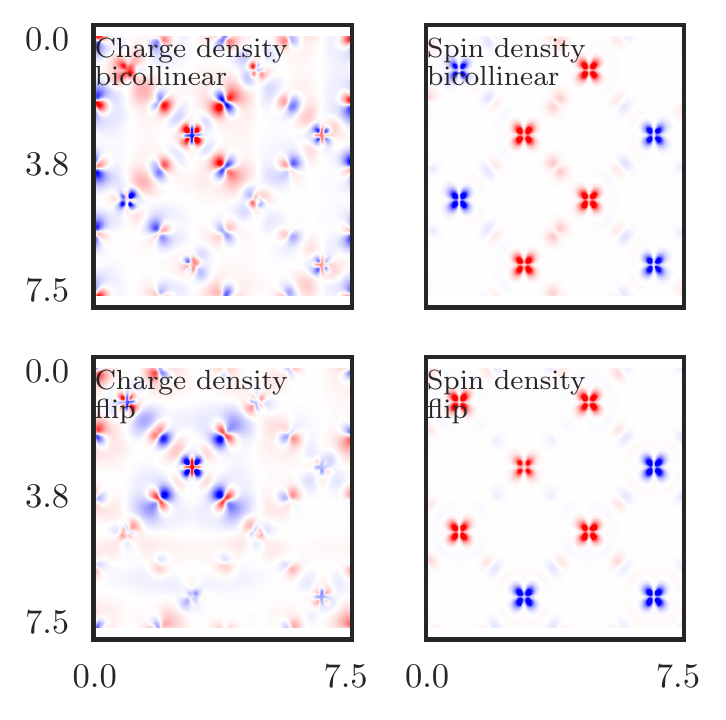}
\end{tabular}

  \caption{\textbf{Spin charge coupling in the Fe-pnictides and cuprates.}  
  The difference of charge density from the average and the total spin density for different spin orders and materials. The $x$ and $y$ axes are $x$ and $y$ in the plane in \AA. 
  Red is positive and blue is negative. 
  While the charge response in these materials is different in nature, they both have a moderate response of the charge to changes in the spin density.}  \label{magconf}
\end{center}
\end{figure}

We use fixed node diffusion Monte Carlo (DMC) in this study. 
This method provides a fully first-principles stochastic framework to address the Schr\"odinger equation and yields a variational upper bound to the ground state~\cite{foulkes2001quantum}. 
We employ a Slater-Jastrow trial wavefunction, as implemented in the {\sc qwalk} package~\cite{wagner2009qwalk}. 
We construct the Slater determinant with orbitals from density functional theory (DFT) calculations using {\sc crystal} code~\cite{dovesi2014crystal14}, employing the PBE0 functional~\cite{perdew1996rationale}.
We use Dirac-Fock pseudopotentials specially constructed for quantum Monte Carlo computations~\cite{burkatzki2007energy,burkatzki2008energy}. 
This combination has been demonstrated to give excellent description of a number of challenging magnetic materials~\cite{0034-4885-79-9-094501}, including cuprates~\cite{wagner2015ground} and FeSe~\cite{busemeyer2016competing}. 
We control finite-size errors using $2\times 2\times 1$ supercells, as well as by averaging over 8 twisted boundary conditions. 
We use a timestep of 0.02 Ha$^{-1}$, while also carrying out checks at a smaller timestep of 0.01 Ha$^{-1}$.
These settings have been shown to have minimal approximations for the Fe-based superconducting materials~\cite{busemeyer2016competing} and cuprates~\cite{wagner2015ground}.
In our study, we consider a number of magnetic configurations, spin densities for some of which are shown in Fig.~\ref{magconf}. 
We investigated collinear, flip (single spin defect in the collinear order), checkerboard, bicollinear orderings, along with ferromagnetic and unpolarized states.

To establish the accuracy of DMC for these 122 materials, we examined the energies of different magnetic configurations for all the compounds (Fig.~\ref{energy}). 
The experimentally known magnetic ground states are the following: BaMn$_2$As$_2$-checkerboard~\cite{singh2009itinerant}, BaCr$_2$As$_2$-checkerboard~\cite{an2009electronic}, BaFe$_2$As$_2$-collinear~\cite{wang2009anisotropy}, BaCo$_2$As$_2$-debated~\cite{anand2014crystallography,tranquada2016}, BaNi$_2$As$_2$-paramagnetic~\cite{pfisterer1983}, and BaCu$_2$As$_2$-paramagnetic~\cite{pfisterer1983}. 
The ground state magnetic order in BaCo$_2$As$_2$ is not clear at present and the neutron scattering data could be consistent with ferromagnetic or collinear order~\cite{anand2014crystallography}, as well as ferrimagnetic order~\cite{tranquada2016}.
BaCr$_2$As$_2$ and BaMn$_2$As$_2$ have checkerboard ground state order.
Most importantly, we obtain collinear magnetic ground state for BaFe$_2$As$_2$, consistent with experimental report~\cite{wang2009anisotropy}. 
Rather surprisingly, for BaCo$_2$As$_2$ we find that none of the the simple magnetic orders have the lowest magnetic energy. 
A defective structure we call `flip' order has the lowest energy, while collinear order is the next lowest in energy.
For BaNi$_2$As$_2$ and BaCu$_2$As$_2$, the unpolarized state has the lowest energy, in agreement with experiments~\cite{pfisterer1983}.
We have also considered doped Ca$_2$CuO$_2$Cl$_2$, whose basic properties using QMC have been described in a previous paper~\cite{wagner2015ground}.

Now we wish to investigate a way of probing whether these materials are described by a model with similar features to the Kondo-Heisenberg model. 
One clear feature, no matter the lattice, is that if the spins are held fixed in this model, then they act as an external potential on the fermions. 
Thus for different fixed spin configurations, one would expect to see a response in the electron density of the ground state. 
We quantify this by measuring the mean absolute variation of the density as a function of the spin order, as outlined in Fig~\ref{magconf}. 
On the set of magnetic configurations, we evaluate the one-particle charge density for each state $\rho_i(r)$ and the spin density $s_i(r)$. 
We form the average spin and charge densities $\bar{\rho}(r) = \frac{1}{N} \sum_i \rho_i(r)$ and $\bar{s}(r) = \frac{1}{N} \sum_i s_i(r)$.

The charge susceptibility to spin for a given state $i$ is computed as 
\begin{equation}
\chi_{i} = \frac{\Delta \rho_i}{\Delta s_i} = \frac{\int |\rho_i(r) - \bar{\rho}(r)| dr}{\int{|s_i(r)- \bar{s}(r)| dr}},
\label{eqn:susceptibility}
\end{equation}
and the average susceptibility is then $\bar{\chi}= \frac{1}{N} \sum_i \chi_i$. 
This is a measure of how much the charge density responds to changes in the spin density. 
Through analogy to the Kondo-Heisenberg model, we would a priori expect that there should be a sweet spot in susceptibility--too little and there is not enough coupling to induce superconductivity, and too much and the system collapses into polaron-like behavior.

\begin{figure*}
\begin{center}
  \includegraphics[scale=1.0]{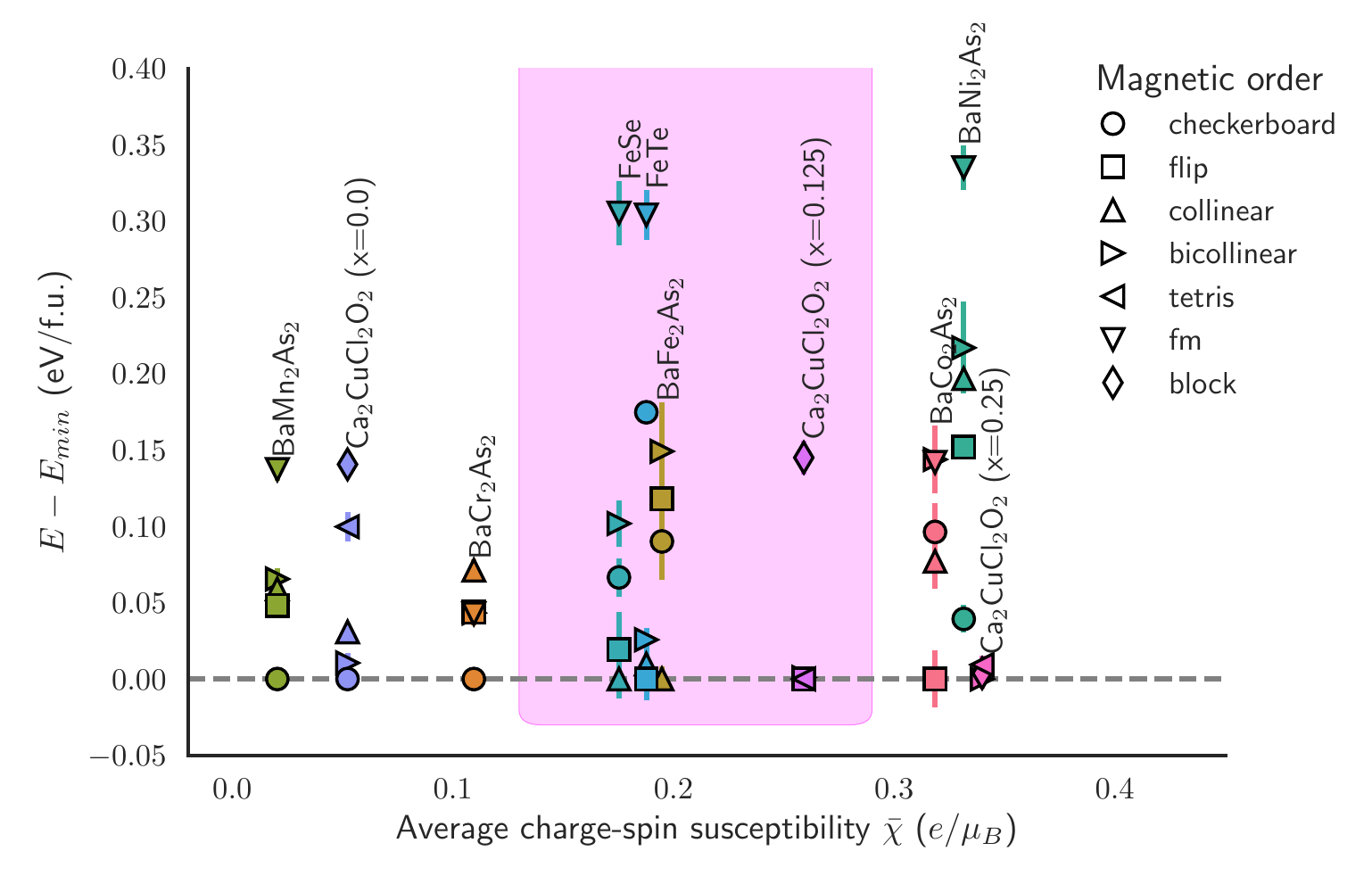}
  \caption{\textbf{A comparison across compounds.}
  Average charge-spin susceptibility versus the relative energies of magnetic orders.
  Symbol shapes indicate different magnetic orders and colors indicate different materials as labeled on the graph. 
  The intermediate coupling region is outlined in pink as a guide to the eye.
  }  \label{acrosscomp}
\end{center}
\end{figure*}

The average charge susceptibility is plotted for a set of superconducting materials and materials similar to superconductors in Fig~\ref{acrosscomp}, along with the relative energies of the ordered magnetic configurations.
For a diverse set of materials with different spin moments and electronic configurations, $\bar{\chi}$ falls into a range between 0 and 0.4 $e/\mu_B$.
This thus appears to be a universal scale across many compounds. 
We also immediately observe that all superconducting materials have intermediate charge susceptibility, in a clearly separated block from other materials. 
This is true both for the iron-based superconductors and the cuprates.
 $\bar{\chi}$ is intermediate for Ca$_2$CuO$_2$Cl$_2$ at 1/8 doping and then is too large with 1/4 doping. 
 This qualitatively parallels the experimental superconducting behavior.

We would like to highlight here the case of BaNi$_2$As$_2$ and BaCo$_2$As$_2$, both of which have large amounts of spin-charge coupling. 
BaNi$_2$As$_2$ has paramagnetic ground state order and becomes a superconductor below 0.7 K~\cite{ronning2008first}. 
If BaNi$_2$As$_2$ could be made to have a magnetic ground state, its superconducting T$_c$ could increase. 
On the other hand, BaCo$_2$As$_2$ has an unusual `flip' ground state which includes a defect in the magnetic order. 
This may be due to the large amount of charge susceptibility the material experiences. 
A modification of this material to decrease the relative effects of the spin on the charge may result in superconductivity. 
This might be done by increasing the bandwidth of the material, perhaps by applying pressure.

\textit{Summary:} In summary, we have shown that diffusion Monte Carlo provides a faithful description of magnetism in transition metal pnictide compounds. 
In particular, the calculated ground states for all compounds in this class are in agreement with experimental reports.
A comparison with more traditionally used density functional theory calculations highlights the importance of including explicit correlations in describing these materials.
Our results show the importance of spin-charge coupling as a key normal state ingredient for high temperature superconductivity in both iron-based and copper-based superconductors. 
Based on an analysis of charge response to magnetic order within the accurate diffusion Monte Carlo framework, we have introduced a spin-charge coupling descriptor that separates superconducting materials from non-superconducting materials. 
Such a calculable descriptor could pave the way for theory-guided search for new unconventional superconductors and gives interesting guidance for modifying known materials to perhaps find superconductivity. 
 
\textit{Acknowledgments:} This work was supported by the Center for Emergent Superconductivity, an Energy Frontier Research Center funded by the U.S. Department of Energy, Office of Science, Office of Basic Energy Sciences under Award Number DEAC0298CH1088. 
We acknowledge illuminating discussions with D. Ceperley, A. Leggett, D. Shoemaker and J. Tranquada. 
Computational resources were provided by the University of Illinois Campus Cluster and the Argonne Leadership Computing Facility.
All data will be available through the Materials Data Facility.

\bibliography{references}

\end{document}